\documentclass[aps,prl,reprint,superscriptaddress]{revtex4-1}

\usepackage{amsmath} 
\usepackage{amssymb}
\usepackage{graphicx} 
\usepackage{float}
\usepackage{color}
\usepackage{listings}
\usepackage{soul}

\newcommand{\figurewidth}{\columnwidth}

\begin{document}

\title{Data-driven learning of total and local energies in elemental boron}
\author{Volker L.\ Deringer}
\email{vld24@cam.ac.uk}
\affiliation{Engineering Laboratory, University of Cambridge,
             Trumpington Street, Cambridge CB2 1PZ, UK}
\affiliation{Department of Chemistry, University of Cambridge,
             Lensfield Road, Cambridge CB2 1EW, UK}
\author{Chris J.\ Pickard}
\affiliation{Department of Materials Science and Metallurgy,
             University of Cambridge,
             27 Charles Babbage Road, Cambridge CB3 0FS, UK}
\affiliation{Advanced Institute for Materials Research,
             Tohoku University 2-1-1 Katahira, Aoba, Sendai, 
             980-8577, Japan}
\author{G\'abor Cs\'anyi}
\affiliation{Engineering Laboratory, University of Cambridge,
             Trumpington Street, Cambridge CB2 1PZ, UK}

\date{\today}

\begin{abstract}
The allotropes of boron continue to challenge 
structural elucidation and solid-state theory.
Here we use machine learning combined with random structure
searching (RSS) algorithms to systematically construct an interatomic
potential for boron.
Starting from ensembles of randomized atomic configurations,
we use alternating single-point quantum-mechanical energy and 
force computations, Gaussian approximation potential
(GAP) fitting, and GAP-driven RSS to iteratively generate a
representation of the element's potential-energy surface.
Beyond the total energies of the very different boron allotropes,
our model readily provides atom-resolved,
local energies and thus deepened insight into the frustrated 
$\beta$-rhombohedral boron structure.
Our results open the door for the efficient and automated generation of
GAPs and other machine-learning-based interatomic potentials, and suggest
their usefulness as a tool for materials discovery.
\end{abstract}

\maketitle

Elemental boron presents a number of complex crystal structures as
a direct consequence of its unique and electron-deficient bonding nature \cite{Albert2009, Oganov2009a, Ogitsu2013}.
This poses formidable challenges for experimentalists and theorists alike. 
Among the most fundamental is determining the thermodynamically stable ground state between two competing forms:
$\alpha$-rhombohedral boron, which contains B$_{12}$ icosahedra exclusively \cite{McCarty1958}, 
and $\beta$-rhombohedral boron ($\beta$-B in the following), which exhibits partial occupations and  geometric frustration, 
most unusual for an elemental ground-state structure \cite{Hoard1970, Callmer1977, Slack1988, Ogitsu2009, Ogitsu2010}.
The energy difference between the $\alpha$ and $\beta$ forms has been studied extensively using 
density-functional theory (DFT) \cite{Prasad2005, Masago2006, VanSetten2007, Widom2008, 
Ogitsu2009, Siberchicot2009, Ogitsu2010, An2016} and recently probed by calorimetric experiments \cite{White2015}; the
consensus is now that $\beta$-B is indeed more stable at ambient conditions.

In recent years, DFT has played a central role not only in understanding $\beta$-B but also in the discovery and structural
elucidation of other allotropes.  Prominently, a high-pressure structure dubbed $\gamma$-B$_{28}$
has been determined with the aid of evolutionary crystal-structure searching
\cite{Oganov2009} as well as direct methods \cite{Zarechnaya2009, Mondal2011}. 
Similar techniques have recently identified ``borophenes'' and other two-dimensional boron allotropes with interesting 
electronic properties \cite{Zhou2014, Mannix2015, Ma2016}. 

Despite their widespread successes \cite{Oganov2011}, DFT-based structure-searching algorithms
are severely restricted by their high computational cost. While excellent empirical interatomic potentials are 
available today for many solid-state systems, there is a conspicuous lack of such potentials for boron \cite{Pokatashkin2015}---the single
interatomic potential in the literature, as the authors stress, is fitted to quantum-mechanical reference data for
$\alpha$-boron exclusively \cite{Pokatashkin2015}. No empirical potential is known to us that could reliably 
describe the potential-energy surface (PES) of elemental boron applicable to multiple phases.

To overcome the performance and scaling limitations of DFT, 
machine-learning (ML) based interatomic potentials are nowadays increasingly used for materials simulations
\cite{[{For a recent overview, see }][]Behler2016}
and have been suggested to be useful for structure searching \cite{Dolgirev2016}. Indeed, we have very recently shown,
as a proof-of-concept, that an ML-based interatomic potential initially fitted for liquid
and amorphous carbon \cite{aC_GAP} can be used to discover hitherto
unknown hypothetical carbon allotropes \cite{C_AIRSS}. However, many carbon (as well as silicon) networks can readily
be generated by direct enumeration \cite{Strong2004, Hoffmann2016, Jantke2017}.
It would now be interesting to apply ML-driven searches to a system with more complex structure and bonding, for which 
boron is an ideal and challenging test case.

In this work, we demonstrate how an ML-based interatomic potential can be systematically constructed by iterative exploration of 
configuration space. It was recently shown that evolutionary searches provide diverse and representative input structures for fitting ML potentials
\cite{Hajinazar2017}, but here we take this concept further by searching ``on the fly''
\cite{LOTF-note, LOTF-1, LOTF-2}: our structure search is driven not by DFT
but by the ML model itself and generates input data for the next iteration of ML training; this is repeated
until a ``consistent'' model is achieved, i.e. one that does not generate new structures that significantly alter the potential when added to the
training set.  Therefore, the searching is neither done {\em ex post}, as in our previous work 
\cite{C_AIRSS}, nor merely as a means to an end,  but here it is the central technique we use for exploring {\em and} fitting a complex
PES.

\begin{figure}[t]
\centering
\includegraphics[width=\figurewidth]{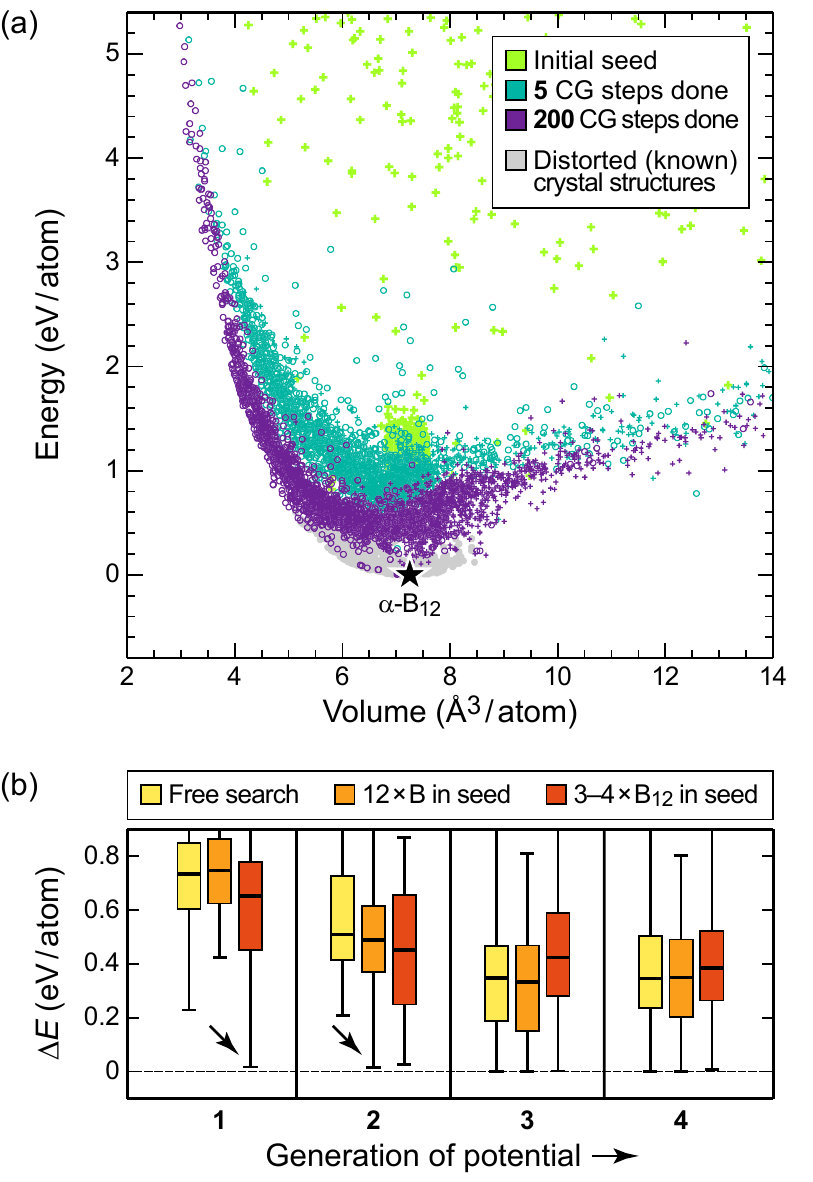}
\caption{\label{fig:EV_landscape}
         Iterative construction of a machine-learning model for the potential-energy
         surface of boron.
         (a) Energy--volume plot for the database of DFT reference computations,
         progressively generated by searching at ambient pressure 
         (crosses) and later at elevated pressure (circles). We start
         with an initial seed of random structures (light green), generated
         in three different ways. We iteratively
         generate new seeds and relax them using the previous version of the GAP,
         adding datapoints after 5 (teal) and 200 (purple) conjugate--gradient
         (CG) steps. All fits furthermore include distorted unit cells of the
         $\alpha$-B$_{12}$, $\gamma$-B$_{28}$, and $\alpha$-Ga type;
         the optimized  
         $\alpha$-B$_{12}$ structure is set as the energy zero.
         (b) DFT energy (as distance from the convex hull) for structures after 
         200 CG steps of GAP-RSS minimization, starting from seeds generated in
         three different ways
         and through consecutive generations of the potential.
         Boxes denote percentiles 
         corresponding to one standard deviation (68\%); bold horizontal lines
         denote the median, and  
         whiskers the entire range of data.}
\end{figure}

We start our search from an ensemble of randomized periodic structures, similar in spirit to the {\em ab initio} random structure
searching (AIRSS) technique \cite{Pickard2006, Pickard2011}. 
Our protocol includes three components, with progressively higher structural ordering and therefore more selective sampling of configuration space. 
First, we generate random structures with 2--32 individual atoms per unit cell
corresponding to a wide range of densities, aiming for a comprehensive sampling of the 
PES including higher-energy regions. Second, we generate structures with 12 at./cell, and search parameters
that would be used to find $\alpha$-B$_{12}$ in an unconstrained DFT-based search. This particular elemental ground state, from our experience, is 
a challenging case for AIRSS and therefore serves as a diagnostic here:
we verified that our method can ``find'' $\alpha$-B$_{12}$ as well. Finally, to sample the PES more accurately around local minima, we build
random structures from a B$_{12}$ icosahedron that is repeated by space-group
symmetry operations (using 36 or 48 at./cell). This is because the B$_{12}$ unit is the most characteristic
building block in boron allotropes \cite{Albert2009}.

We generated 500 random structures with each of these three approaches
and used single-point DFT-PBE computations \cite{Perdew1996} to obtain their energies and forces.
These reference data were generated using dense spacing of $k$ meshes
(0.02 \AA{}$^{-1}$), a plane-wave cutoff energy of 800 eV, and on-the-fly pseudopotentials as implemented
in CASTEP 8.0 \cite{CASTEP}. The resulting energies are shown in Fig.\ \ref{fig:EV_landscape}a as light green points.

The first (and most unconstrained) approach leads to highly scattered
results, as expected, and the unrelaxed trial structures have a median energy of 6.1 eV/at. above $\alpha$-B$_{12}$.
The other seeding procedures give 1.2 and 0.9 eV/at., respectively, and the resulting structures lie much closer together in the
$E$--$V$ plot (Fig. \ref{fig:EV_landscape}a). Normally, one would now relax these structures using DFT 
\cite{Pickard2011, Hajinazar2017}. Instead, we show in the following how 
single-point computations suffice to initialize
a model that, then, progressively explores lower-lying configurations.

To ``machine-learn'' the PES spanned by these seed points, we fitted a Gaussian
Approximation Potential (GAP) model \cite{Bartok2010} with combined two-, three-, and many-body descriptors and a radial cutoff of 3.7 \AA{},
similar to previous work \cite{aC_GAP}. For many-body interactions, we use the Smooth Overlap of Atomic Environments (SOAP) kernel 
\cite{Bartok2013}. The fit also includes distorted unit cells of low-lying
crystalline polymorphs (gray in Fig.\ \ref{fig:EV_landscape}a), which we found to be required for finding the $\alpha$-B$_{12}$ ground-state
structure using early generations of the GAP.

With the initial potential available, we performed random structure searching (denoted as GAP-RSS in the following), 
creating new seeds as above but now relaxing them using GAP. We took snapshots of these relaxations after 5
and 200 conjugate--gradient steps, respectively: the former sample the early stages of GAP-RSS relaxation trajectories; the latter are close to
the local minima. Results of single-point DFT computations on these snapshots were added to the database
after each iteration and are included in Fig.\ \ref{fig:EV_landscape}a (teal
and purple symbols, respectively). We performed this process twice,
consecutively, at ambient pressure (crosses in Fig.\ \ref{fig:EV_landscape}a; potential
generation {\bf 1--2}). Then, we repeated the search two more times at elevated
pressure values, drawn from an exponential distribution with a width of 100 GPa (circles in Fig.\ \ref{fig:EV_landscape}a;
potential generation {\bf 3--4}). This serves to sample more densely packed structures.

The evolution of the training database is illustrated in Fig.\ \ref{fig:EV_landscape}b, focusing on the
energy of the relaxed structures from our RSS iterations. The results are given as distance from the convex hull, 
and separately for the three seeding procedures described above. Clearly, the more ordered
seeds initially produce lower-energy structures (arrows), but this changes with growing quality of the potential. In generation
{\bf 4}, all three components of our protocol led to comparable energies after 
relaxation of the corresponding seed structures
(rightmost panel in Fig.\ \ref{fig:EV_landscape}b). 
The final training database was extended by adding distorted $\beta$-B
structures, which were not required during iterative fitting but
needed to accurately describe the energy of the $\beta$ form.
It contains the results of 8,388 single-point
computations (of which 3,000 are end points of GAP-RSS minimizations), 
corresponding to over 210,000 individual atomic environments. We emphasize that only single point DFT calculations were performed;
no sampling or searching was done with DFT.

\begin{figure}[t]
\centering
\includegraphics[width=\figurewidth]{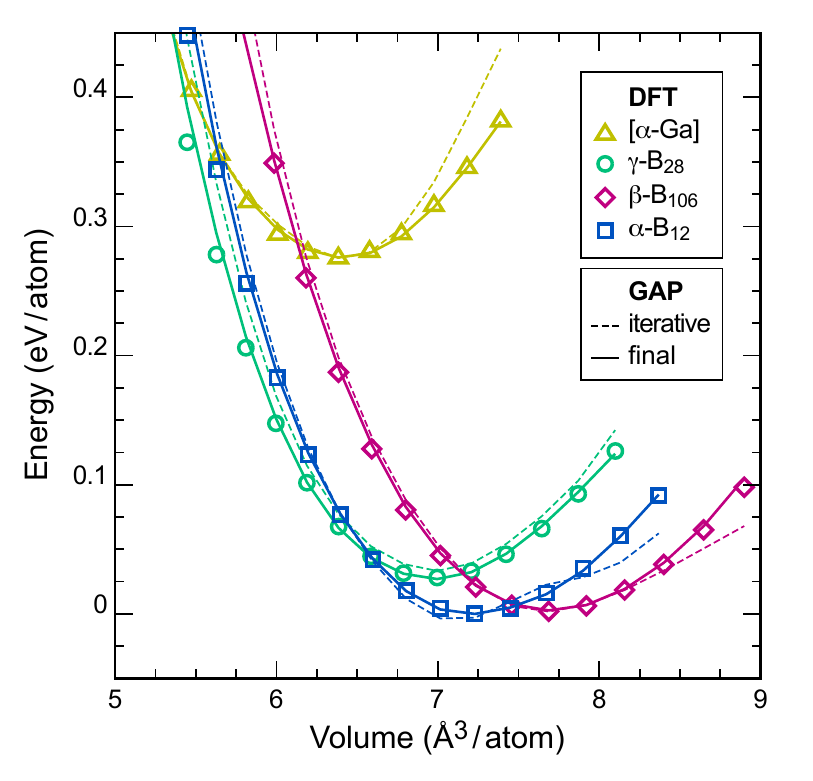}
\caption{\label{fig:EVs_vs_DFT}
         Energy--volume plots for relevant boron
         allotropes. Symbols denote results of DFT computations,
         whereas lines connect GAP energies for the same structures.
         Two versions of the potential have been fitted to the complete
         reference database: one (thin dashed lines) using the smooth
         settings as during the iterative procedure, and one (thick
         lines) with the final, tighter settings.}
\end{figure}

We found that the results of our procedure depend strongly on 
the smoothness of the potential, which is controlled by a single parameter, $\sigma_{\rm at}$,
in the SOAP formalism \cite{Bartok2013}. A setting of $\sigma_{\rm at} = 0.5$ \AA{} was previously used successfully to
fit GAP models for carbon \cite{aC_GAP} and tungsten \cite{Szlachta2014},
but did not produce stable potentials for early-generation GAP-RSS minimizations in our
experiments.
A smoother potential is therefore required to interpolate between the high-energy datapoints
at the early stages of GAP-RSS, while still being accurate enough for finding
local minima at all. We found $\sigma_{\rm at} = 0.75$ \AA{} to be a viable choice for structure
searching and used this throughout the iterations, together with 
$n_{\rm max} = l_{\rm max} = 8$ for the spherical-harmonics expansion
of the neighbor density in SOAP (as described in Ref.\ \cite{Bartok2013}). 
Once the database was completed, we performed a final fit on the same 
database but with tighter settings of $\sigma_{\rm at} = 0.5$ \AA{} and
$n_{\rm max} = l_{\rm max} = 12$. 

To validate our potential, we computed DFT energies at varied unit-cell volumes
for the most important boron allotropes \cite{Oganov2009}, 
viz.\ $\alpha$-B$_{12}$, $\gamma$-B$_{12}$,
a high-pressure $\alpha$-Ga type polymorph, and $\beta$-B. To represent
the disordered structure of the latter, we use the discrete
``$\beta$-B$_{106}$'' structural model (see below)
which provides a good approximant of the experimentally determined mixed occupations
and is practically degenerate with $\alpha$-B$_{12}$ in DFT energy \cite{VanSetten2007}.
Indeed, Fig.\ \ref{fig:EVs_vs_DFT} illustrates immediately why the PES
of boron is such a challenging case: the energy differences between
three polymorphs containing B$_{12}$ units are tiny, and only the
fundamentally different [$\alpha$-Ga] structure (not containing any
B$_{12}$ icosahedra) is clearly distinct from the other polymorphs on this energy scale. Still,
our final GAP reproduces the energy of all these structures extremely
well (Fig.\ \ref{fig:EVs_vs_DFT}).

\begin{figure}[t]
\centering
\includegraphics[width=\figurewidth]{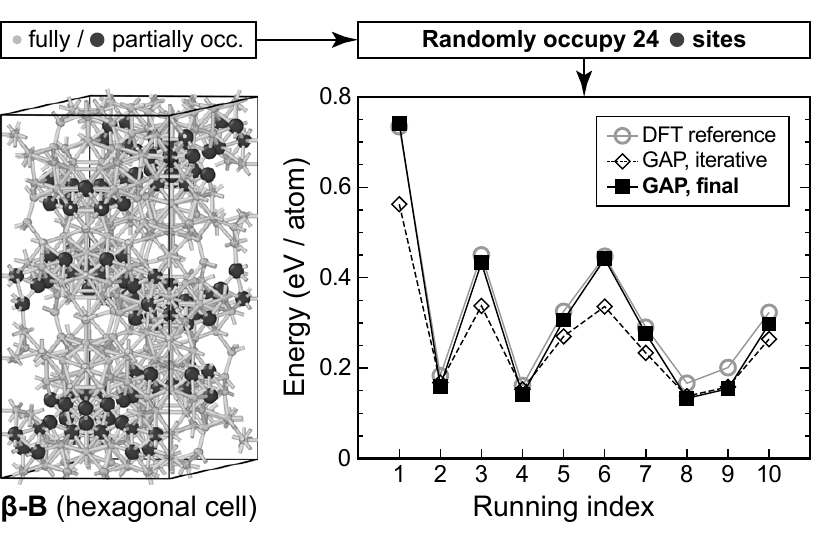}
\caption{\label{fig:compare_B321}
         Stability of randomized mixed-occupation models for $\beta$-B
         as a test for the transferability of the potential.
         The structural drawing shows the hexagonal unit-cell setup
         according to Ref.\ \cite{Slack1988}: fully occupied Wyckoff 
         sites are
         indicated by small, light gray atoms, whereas partially
         occupied sites are shown by larger, black atoms. On the latter
         sites, we randomly distribute 24 B atoms, which, together with
         297 atoms on fully occupied sites, constitute the $\beta$-B$_{321}$
         model. The right-hand side shows energies for ten arbitrary,
         discrete structural models created in this fashion.}
\end{figure}

As a further test, we generated discrete trial structures
with site occupations that are not part of the training
(Fig.\ \ref{fig:compare_B321}): recall that
our reference database does contain $\beta$-B,
but describes it using only one
particular set of occupations ($\beta$-B$_{106}$). Here,
by contrast, we started from the most detailed structural model available,
which was proposed based on single-crystal
X-ray diffraction from a highly pure sample (carbon impurities $\approx 150$ ppm) \cite{Slack1988}. Besides the known B13 and B16
sites (which have previously reported site occupations of approximately $3/4$ and $1/4$, respectively; Ref.\ \cite{Callmer1977}), this model
contains additional partially occupied sites, labeled B17--B20, with occupations of just a few percent \cite{Slack1988}. We randomly generated
ten structural models, none of which had been included in the GAP fitting, and compared their DFT and GAP energies. Again, the potential fitted with
GAP-RSS settings (``iterative'') already captures all qualitative trends, but re-fitting with tighter settings (``final'') 
is needed to bring the results close to quantitative agreement. Clearly, different smoothness requirements exist for
potentials that (i) explore and (ii) quantify a PES.

We finally show how a common feature of interatomic potentials
can here be turned into a distinct advantage. Interatomic potentials for materials, by their essential nature and similarly to 
biomolecular force fields, are typically a combination of  terms for long-range interactions (describing electrostatics and dispersion) and
short-range (often called ``bonded'') interactions \cite{Behler2016}. We focus here on the latter, which can be understood as a decomposition the
the total energy $E$ of a collection of atoms into a sum of atomic contributions,
\[
 E = \sum_{i} \varepsilon_{i}
\]
where the local atomic energy  
$\varepsilon_{i}$ depends only on a region around the $i$-th atom as
specified by a cutoff radius (here, $r_{\rm cut} = 3.7$ \AA{}).
While this intrinsically limits the attainable accuracy of the potential 
(since quantum mechanics is fundamentally long-ranged) \cite{Bartok2010, aC_GAP},
this approximation is often rather good, 
and here it allows us to analyze the local stability of atoms---which otherwise is not straightforwardly possible within a quantum-mechanical
framework. This is particularly interesting for $\beta$-B with its partly occupied sites.  

\begin{figure}[t]
\centering
\includegraphics[width=\figurewidth]{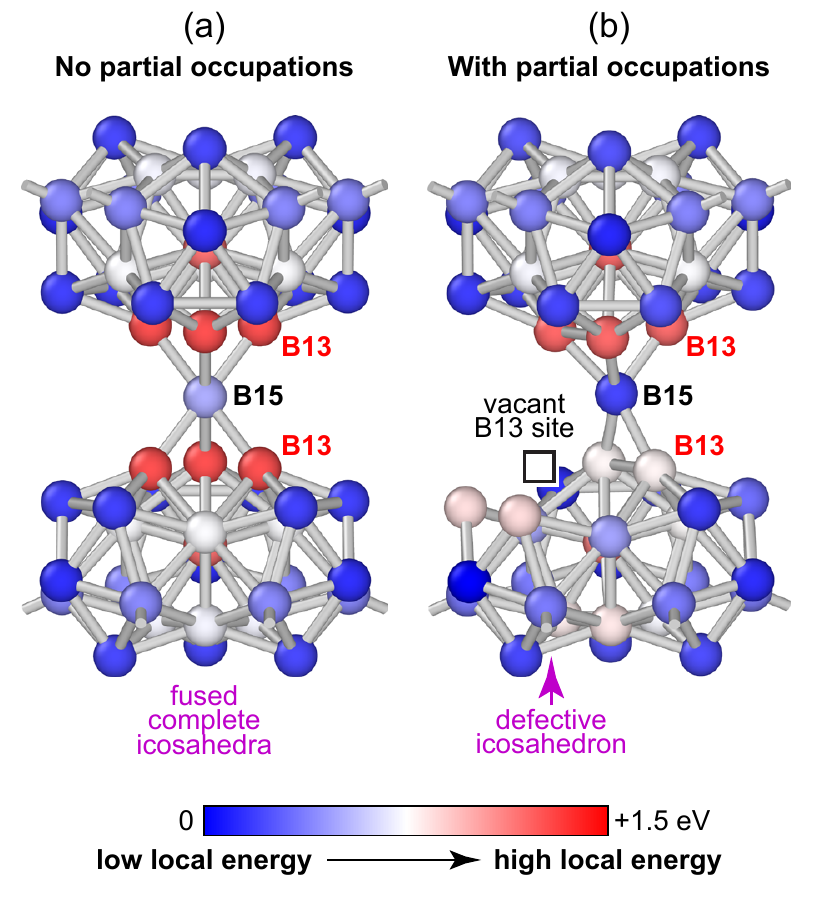}
\caption{\label{fig:local_heat}
         Local energy analysis using our GAP model, comparing the simplistic
         $\beta$-B$_{105}$ model without partial occupations (a) to the more
         disordered and favorable $\beta$-B$_{106}$ model (b). The characteristic
         structural fragment at the center of the $\beta$-B unit cell
         is shown, with atoms color-coded according to more favorable
         (blue) to less favorable (red) environments. 
         The labeling of crystallographic sites is as in the original
         literature; the analysis was performed for optimized structures.
         Visualization was
         done using OVITO \cite{OVITO}.}
\end{figure}

Figure \ref{fig:local_heat}a shows GAP-computed local energies $\varepsilon_{i}$  for individual atoms in a characteristic
fragment from $\beta$-B, described using the simplistic $\beta$-B$_{105}$ model without any partial occupations 
\cite{Geist1970}. Besides icosahedral B$_{12}$ units, this structure contains B$_{28}$ building blocks that can be regarded
as triply fused icosahedra.  Figure \ref{fig:local_heat}a shows two such complete clusters, 
connected to an isolated atom (B15) at the center of the unit cell, via three apical atoms (B13) each.
However, the latter site has been shown to be only partially occupied \cite{Hoard1970, Callmer1977, Slack1988},
and the GAP analysis corroborates this: a full occupation of the B13 sites is clearly unfavorable
due to high local energies (red).

Indeed, a more favorable structural model is obtained when only five of six B13 sites are occupied (Fig.\ \ref{fig:local_heat}b)
\cite{VanSetten2007}, in accord with the more recent structural refinements in the experimental literature \cite{Hoard1970, Callmer1977, Slack1988}.
In this case, which corresponds to the $\beta$-B$_{106}$ model used in Fig.\ 2, the local energies of the remaining two B13 atoms are significantly lowered. Vacancy
formation also stabilizes the central B15 atom. In turn, one of the constituent icosahedra is now defective
due to the presence of a vacancy ($\square$), and therefore the neighboring atoms have higher local energies (pale red) than those in the complete
fragment above (dark blue).

In conclusion, we have generated an interatomic potential for elemental boron that describes the energetics of multiple polymorphs, using a machine-learning approach based on DFT input data.
Our protocol requires single-point DFT calculations only, and therefore can explore complex configuration spaces with reasonable
computational effort---boron is well described using the economic PBE functional \cite{Oganov2009},
but other materials will require higher-level data such as hybrid DFT, where
relaxations of thousands of structures are not readily possible.
We used the Gaussian Approximation Potential framework for generating the model, but other methods such as artificial neural networks can be combined with the same ideas.
In the future, coupling machine learning with random structure searching may enable the on-the-fly
construction of interatomic potentials of unprecedented level of quality for the purpose of materials discovery.\\[-4mm]

\begin{acknowledgments}

We thank N.\ Bernstein, S.\ R.\ Elliott, and D.\ M.\ Proserpio
for ongoing valuable discussions and collaborations.
V.L.D.\ gratefully acknowledges a Feodor Lynen fellowship
from the Alexander von Humboldt Foundation, a Leverhulme
Early Career Fellowship, and
support from the Isaac Newton Trust (Trinity College, Cambridge). 
C.J.P. is supported by the Royal Society through
a Royal Society Wolfson Research Merit award.
This work used the ARCHER UK National Supercomputing
Service (http://www.archer.ac.uk) via 
EPSRC grants EP/K014560/1 and EP/P022596/1.

{\em Data access statement:} Original data supporting this
publication, including DFT data and GAP parameter
files, will be made freely available in an online repository
upon acceptance.

\end{acknowledgments}


\begin{thebibliography}{35}
\bibitem{LOTF-note}
This is conceptually reminiscent of, but different in detail from,
recently proposed schemes to ``machine-learn'' forces on the fly during
molecular-dynamics simulations \cite{LOTF-1, LOTF-2}.
\end{thebibliography}
\end{document}